\documentclass[epj]{webofc}
\usepackage[utf8]{inputenc}
\usepackage[varg]{txfonts}   
\usepackage{booktabs}
\usepackage{xcolor}

\newcommand{\lapprox}{\raisebox{-0.5ex}{$\
\stackrel{\textstyle<}{\textstyle\sim}\ $}}
\definecolor{darkred}{rgb}{0.4,0.0,0.0}
\definecolor{darkgreen}{rgb}{0.0,0.4,0.0}
\definecolor{darkblue}{rgb}{0.0,0.0,0.4}
\usepackage[bookmarks,linktocpage,colorlinks,
    linkcolor = darkred,
    urlcolor  = darkblue,
    citecolor = darkgreen]{hyperref}
\usepackage{subfigure}
\wocname{EPJ Web of Conferences}
\woctitle{Lattice2017}
\begin{document}
%
\selectlanguage{english}
\title{%
Numerical study of the $2+1d$ Thirring model\\ with U($2N$)-invariant fermions
}
\author{%
\firstname{Simon} \lastname{Hands}\inst{1}\fnsep\thanks{Acknowledges financial support
from  STFC and the Leverhulme Trust
}}
\institute{%
Department of Physics, College of Science,\\
Swansea University, Singleton Park, Swansea SA2 8PP, United Kingdom}
\abstract{%
In 2+1 dimensions the global U($2N$) symmetry associated with massless Dirac fermions is
broken to U($N)\otimes$U($N$) by a parity-invariant mass. I will show how to adapt the
domain wall formulation to recover the U($2N$)-invariant limit in interacting
fermion models as the domain wall separation is increased. In particular, I will
focus on the issue of potential dynamical mass generation in the Thirring model,
postulated to take place for $N$ less than some critical $N_c$. I will present
results of simulations of the model using both HMC ($N=2$) and RHMC ($N=1$)
algorithms, and show that the outcome is very different from previous numerical
studies of the model made with staggered fermions, where the corresponding
pattern of symmetry breaking is distinct.
}
\maketitle
\section{Introduction}\label{intro}

Relativistic fermions moving in 2+1 spacetime dimensions form
the basis of some facscinating quantum field theories, whose principal 
applications mainly fall in the domain of the condensed matter physics of layered
systems. For example, fermion degrees of freedom described by a
Dirac equation have been invoked and studied in models of the underdoped regime
of $d$-wave superconductors~\cite{Tesanovic:2002zz, Herbut:2002yq}, 
where they arise as excitations at the nodes of the
gap function $\Delta(\vec k)$; in models of spin-liquid behaviour in Heisenberg
antiferromagnets~\cite{Wen:2002zz, Rantner:2002zz}, 
where the infrared behaviour of compact QED$_3$ remains an
open issue; as surface states of three-dimensional topological insulators; and of
course as low-energy electronic excitations in graphene~\cite{CastroNeto:2009zz}.

What all these models share in common is their use of four-component {\it
reducible\/} spinor representations for the fermi fields. For non-interacting 
fermions the action in Euclidian metric is 
\begin{equation}
S=\int d^3x\,\bar\psi(\gamma_\mu\partial_\mu)\psi+m\bar\psi\psi;\;\;\mu=0,1,2;
\label{eq:action}
\end{equation}
a key point is the mass term proportional to $m$ is hermitian and invariant
under the parity inversion $x_\mu\mapsto-x_\mu$. For $m=0$
$S$ has a global U(2) invariance generated by 
\begin{equation}
\psi\mapsto e^{i\alpha}\psi,\; \bar\psi\mapsto\bar\psi e^{-i\alpha};\;\;\;\;\;\;
\psi\mapsto e^{\alpha\gamma_3\gamma_5}\psi,\; \bar\psi\mapsto\bar\psi
e^{-\alpha\gamma_3\gamma_5};\label{eq:oneg35}
\end{equation}
\begin{equation}
\psi\mapsto e^{i\alpha\gamma_3}\psi,\; \bar\psi\mapsto\bar\psi e^{i\alpha\gamma_3};\;\;\;
\psi\mapsto e^{i\alpha\gamma_5}\psi,\; \bar\psi\mapsto\bar\psi
e^{i\alpha\gamma_5}.\label{eq:g3g5}
\end{equation}
For $m\not=0$ the $\gamma_3$ and $\gamma_5$ rotations (\ref{eq:g3g5}) are no
longer symmetries and the general pattern of breaking is thus
U($2N)\to$U($N)\otimes$U($N$), where we generalise to $N$ degenerate flavors.

Because there is no chiral anomaly in 2+1$d$, it is possible to perform a change
of variables in the path integral to identify two further ``twisted'' mass terms which,
though antihermitian, are physically equivalent to the $m\bar\psi\psi$  of (\ref{eq:action}):
\begin{equation}
im_3\bar\psi\gamma_3\psi;\;\;\;im_5\bar\psi\gamma_5\psi.
\label{eq:twisted}
\end{equation}
The ``Haldane'' mass term $m_{35}\bar\psi\gamma_3\gamma_5\psi$ is not
equivalent because it changes sign under parity, and will not be considered
further.

\section{The Thirring model in 2+1$d$}\label{Thirring}

A model of particular interest is the Thirring model, which has a contact interaction
between conserved fermion currents. Its Lagrangian density reads
\begin{equation}
{\cal
L}=\bar\psi_i(\partial{\!\!\!/\,}+m)\psi_i+{g^2\over{2N}}(\bar\psi_i\gamma_\mu\psi_i)^2;\;\;i=1,\ldots,N.
\end{equation}
Equivalently, its dynamics is captured by a bosonised action via the introduction
of an auxiliary vector field $A_\mu$:
\begin{equation}
{\cal
L}_A=\bar\psi_i(\partial{\!\!\!/\,}+i{g\over\surd
N}A_\mu\gamma_\mu+m)\psi_i+{1\over2}A_\mu^2.
\label{eq:Thirbose}
\end{equation}
The Thirring model is arguably the simplest interacting theory of fermions
requiring a computational approach. The coupling $g^2$ has dimension $2-d$,
where $d$ is the dimension of spacetime, so a naive expansion in powers of $g^2$
is non-renormalisable for $d>2$. However, things look different after
resummation. First introduce an additional St\"uckelberg scalar $\varphi$ so
the bosonic term becomes ${1\over2}(A_\mu-\partial_\mu\varphi)^2$, to
identify a hidden local symmetry~\cite{Itoh:1994cr}
\begin{equation}
\psi\mapsto e^{i\alpha}\psi;\;\;A_\mu\mapsto
A_\mu+\partial_\mu\alpha;\;\;\varphi\mapsto\varphi+\alpha.
\label{eq:gt}
\end{equation}
This point of view strongly suggests the identification of $A_\mu$ as an abelian
gauge field; the original Thirring model (\ref{eq:Thirbose}) corresponds to a
unitary gauge $\varphi=0$. In Feynman gauge, to leading order in $1/N$ the
resummed vector propagator is now of the form $\langle
A_\mu(k)A_\nu(-k)\rangle\propto\delta_{\mu\nu}/k^{d-2}$. An expansion in powers
of $1/N$ may now be developed by analogy with QED$_3$, and is exactly
renormalisable for $2<d<4$~\cite{Hands:1994kb}.

The outstanding theoretical issue is whether, for $g^2$ sufficiently large and
$N$ sufficiently small, there is a symmetry-breaking transition leading to
formation of a bilinear condensate $\langle\bar\psi\psi\rangle\not=0$
accompanied by dynamical fermion mass generation. 
It is of particular
interest to identify the critical $N_c$ above which symmetry breaking does not
occur even in the strong coupling limit. 
The transitions at
$g_c^2(N<N_c)$ then potentially define a series of distinct quantum critical
points (QCPs). One early prediction, using strong-coupling Schwinger-Dyson
equations in the ladder approximation, found $N_c\simeq4.32$~\cite{Itoh:1994cr}.

\section{The Thirring model with staggered fermions}\label{staggered}

The Thirring model has been studied by numerical simulations using staggered
fermions, with action~\cite{DelDebbio:1997dv}
\begin{equation}
S_{stag}={1\over2}\sum_{x\mu i}\left[\bar\chi_x^i\eta_{\mu x}(1+iA_{\mu
x})\chi_{x+\hat\mu}^i -
\mbox{h.c}\right]+m\sum_{xi}\bar\chi_x^i\chi_x^i+{N\over{4g^2}}A_{\mu x}^2.
\label{eq:Sstag}
\end{equation}
Eq.~(\ref{eq:Sstag}) is not unique, but has the feature that the linear coupling
of the auxiliary precludes higher-point interactions between
fermions once it is integrated over. The action (\ref{eq:Sstag}) has a global
U($N)\otimes$U($N$) symmetry broken to U($N$) by either explicit
or spontaneous mass generation. In a weakly coupled long-wavelength limit a
U($2N_f$) symmetry is recovered with $N_f=2N$~\cite{Burden:1986by}; however the putative QCP is not
weakly coupled, so this conventional wisdom must be questioned. 

\vspace{-0.25cm}
\begin{figure}[thb] 
  \centering
  \includegraphics[width=9cm,clip]{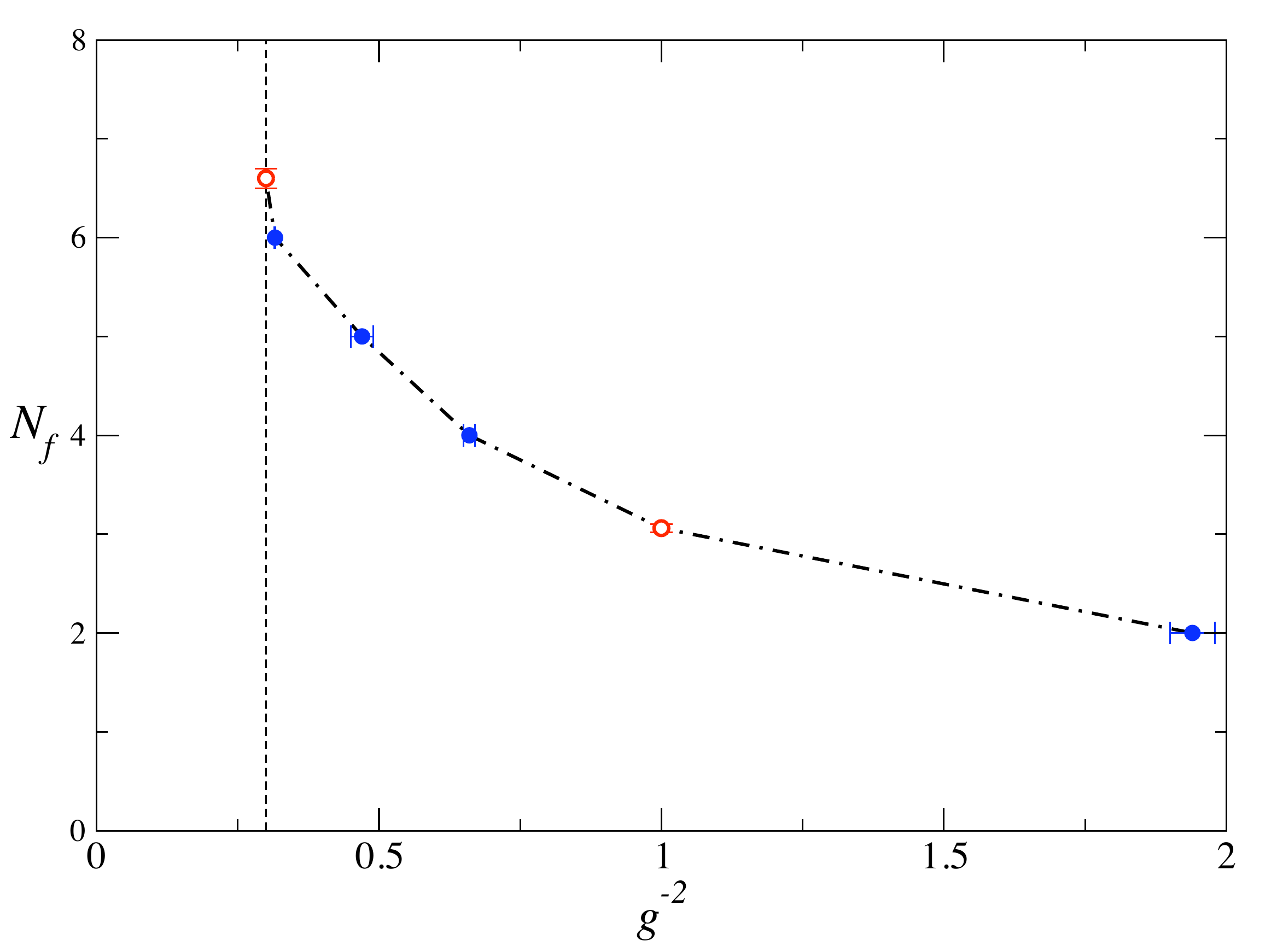}
\vspace{-0.25cm}
  \caption{Phase diagram for the Thirring model with staggered fermions. A
bilinear condensate is spontaneously generated in the region at lower left.
The vertical dashed line denotes the effective strong coupling limit.}
  \label{fig:phased}
\end{figure}
The action (\ref{eq:Sstag}) has been studied for
$N_f\in[2,6]$~\cite{DelDebbio:1997dv,DelDebbio:1999he,Hands:1999id} and QCPs have
been identified and characterised. Other studies with $N_f=2$ have found
compatible results~\cite{Focht:1995ie,Chandrasekharan:2011mn}. 
A more recent study which plausibly identifies the
strong-coupling limit reported $N_{fc}=6.6(1)$, $\delta(N_{fc})\approx7$~\cite{Christofi:2007ye}. 
The results are summarised in
Fig.~\ref{fig:phased}. The ``chiral'' symmetry is indeed broken for small $N_f$ and large
$g^2$, and the exponent $\delta$ defined by the critical scaling
$\langle\bar\chi\chi\rangle\vert_{g_c^2}\propto m^{1\over\delta}$ is very
sensitive to $N_f$. Other exponents can be estimated using hyperscaling.
These results are in qualitative but not
quantitative agreement with the Schwinger-Dyson approach, which predicts
$\delta(N_{fc})=1$. 
A non-covariant form of (\ref{eq:Sstag}) has been used to model the
semimetal-insulator transition in graphene, finding
$N_{fc}\approx5$~\cite{Hands:2008id}
suggesting that a Mott insulating phase is
possible for the $N_f=2$ appropriate for monolayer
graphene~\cite{Armour:2009vj}.

The staggered Thirring model thus exhibits a non-trivial phase diagram, with a
sequence of QCPs with an $N$-dependence  
quite distinct from those of the
theoretically much better-understood Gross-Neveu (GN) model. However, recent
simulations of $N_f=2$ with a fermion bag algorithm which permits study directly
in the massless limit, have found compatible exponents for the 
QCP~\cite{Chandrasekharan:2011mn,Chandrasekharan:2013aya}:
\begin{eqnarray}
\nu&=&0.85(1);\;\;\;\;\,\eta=0.65(1);\;\;\;\;\;\eta_\psi=0.37(1);\;\;\;\;\,\mbox{Thirring}\nonumber\\
\nu&=&0.849(8);\;\;\;\eta=0.633(8);\;\;\;\eta_\psi=0.373(3);\;\;\;\mbox{GN}
\label{eq:exponents}
\end{eqnarray}
The large-$N$ GN values are $\nu=\eta=1$. These results are troubling: from
the perspective of the large-$N$ expansion using bosonised actions the models
should be distinct, whereas (\ref{eq:exponents}) suggest rather they lie in the
same RG basin of attraction. Indeed, when written purely in
terms of four-point interactions between staggered fields spread over elementary
cubes, the only difference between the models is an extra body-diagonal coupling
in the GN case~\cite{Chandrasekharan:2013aya}.  

In this study we examine the possibility that staggered
fermions do not reproduce the expected physics because of a failure to
capture the correct continuum global symmetries near a QCP. A similar insight has
been offered by the Jena group~\cite{Schmidt:2015fps}. 

\section{Domain wall fermions in 2+1$d$}\label{sec:DWF}

The physical idea of domain wall fermions (DWF) is that fermions $\Psi,\bar\Psi$ are allowed to
propagate along an extra fictitious dimension of extent $L_s$ with open boundary
conditions. In 2+1+1$d$ this propagation is governed by an operator
$\sim\partial_3\gamma_3$. As $L_s\to\infty$ zero modes of $D_{DWF}$ localised on the domain walls at
either end become $\pm$ eigenmodes of $\gamma_3$, and physical fields in the
target 2+1$d$ space identified via
\begin{equation}
\psi(x)=P_-\Psi(x,1)+P_+\Psi(x,L_s);\;\;\;
\bar\psi(x)=\bar\Psi(x,L_s)P_-+\bar\Psi(x,1)P_+,
\label{eq:target}
\end{equation}
with $P_\pm={1\over2}(1\pm\gamma_3)$. The walls are then coupled with a term
proportional to the explicit massgap $m$. 

However, the emergence of the U($2N$) symmetry outlined in Sec.~\ref{intro} is
not manifest, because while the wall modes are eigenmodes of $\gamma_3$, the
continuum symmetry (\ref{eq:oneg35},\ref{eq:g3g5}) demands equivalence under
rotations generated by both $\gamma_3$ {\em and} $\gamma_5$. Another way of seeing
this is that the twisted mass terms (\ref{eq:twisted}) should yield identical
physics, eg. the strength of the corresponding bilinear condensate, as
$L_s\to\infty$. This requirement is apparently non-trivial, since while $m$ and
$m_3$ couple $\Psi$ and $\bar\Psi$ fields on opposite walls, $m_5$ couples
fields on the {\em same} wall.

The recovery of U(2$N$) symmetry as $L_s\to\infty$ was demonstrated numerically
in a study of quenched non-compact QED$_3$ on $24^3\times L_s$ systems, for
a range of couplings $\beta$~\cite{Hands:2015qha}. First define the principal residual $\Delta$
via the imaginary part of the twisted condensate:
\begin{equation}
\Im[i\langle\bar\Psi(1)\gamma_3\Psi(L_s)\rangle]=
-\Im[i\langle\bar\Psi(L_s)\gamma_3\Psi(1)\rangle]\equiv\Delta(L_s).
\end{equation}
The difference between the various condensates and their value
in the large-$L_s$ limit is then specified in terms of secondary residuals
$\varepsilon_i(L_s)$ via
\begin{eqnarray}
\langle\bar\psi\psi\rangle_{L_s}&=&i\langle\bar\psi\gamma_3\psi\rangle_{L_s\to\infty}+2\Delta(L_s)
+2\varepsilon_h(L_s);\nonumber\\
i\langle\bar\psi\gamma_3\psi\rangle_{L_s}&=&i\langle\bar\psi\gamma_3\psi\rangle_{L_s\to\infty}
+2\varepsilon_3(L_s);\nonumber\\
i\langle\bar\psi\gamma_5\psi\rangle_{L_s}&=&i\langle\bar\psi\gamma_3\psi\rangle_{L_s\to\infty}
+2\varepsilon_5(L_s).
\end{eqnarray}
\begin{figure}
\begin{minipage}[t]{70mm}
\begin{center}
\includegraphics[width=7.5cm]{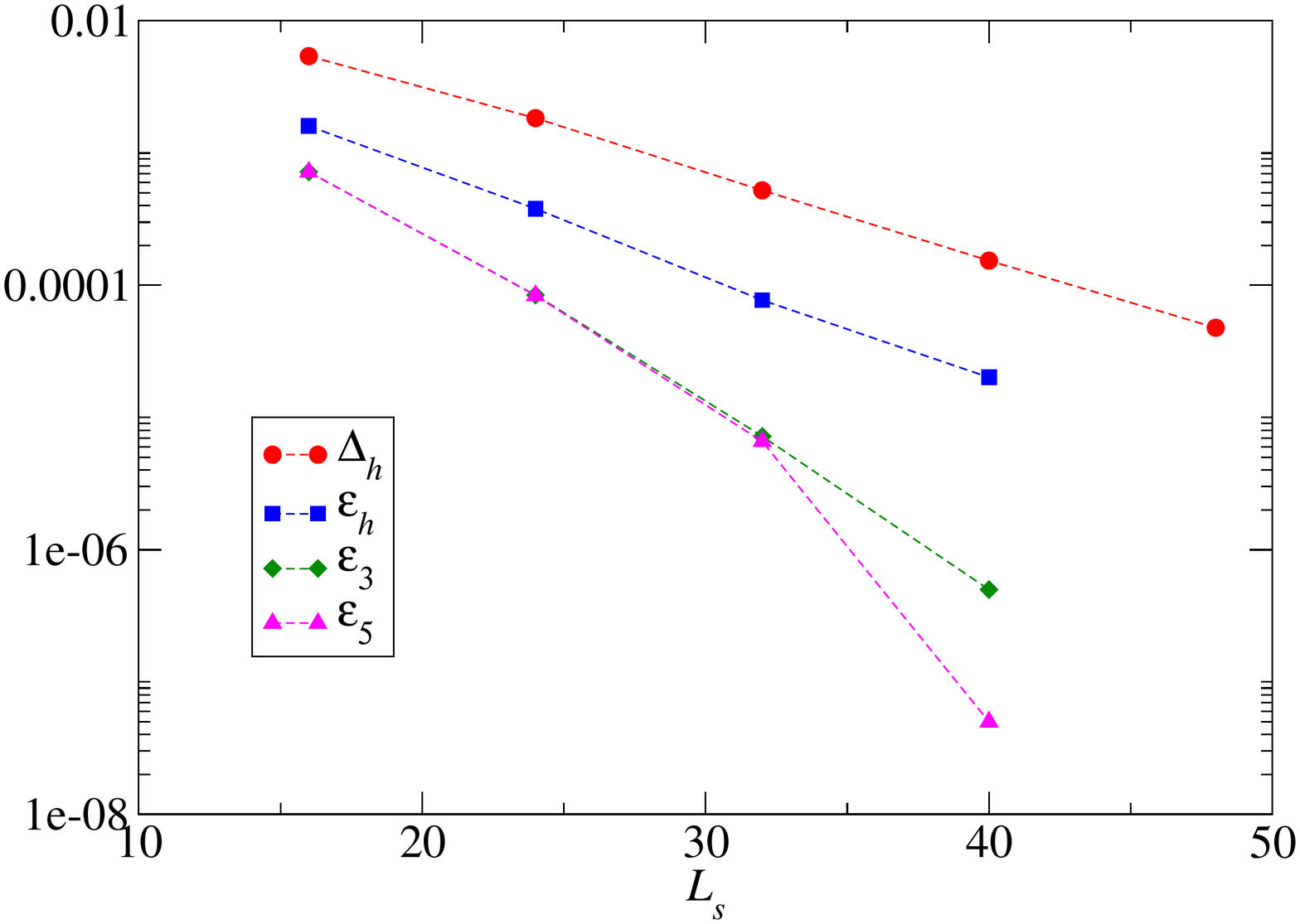}
\caption{Residuals as a function of $L_s$ in \hfill\break quenched non-compact QED$_3$.}
\label{fig:Delta_Ls}
\end{center}
\end{minipage}
\hspace{\fill}
\begin{minipage}[t]{70mm}
\begin{center}
\includegraphics[width=7.5cm]{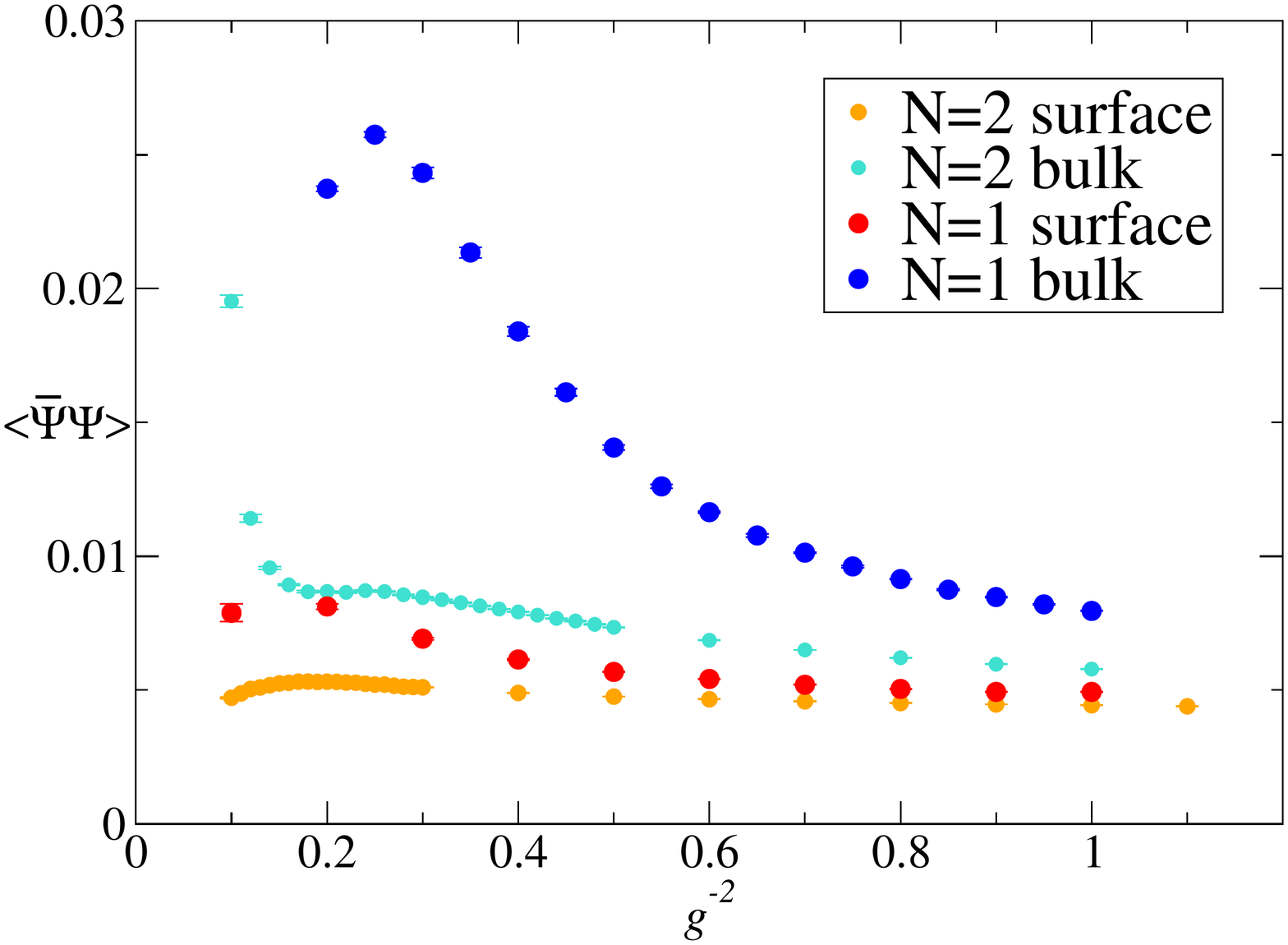}
\caption{Bilinear condensate vs. $g^{-2}$ for both surface and bulk Thirring
models.}
\label{fig:cond_N}
\end{center}
\end{minipage}
\end{figure}
Empirically, as shown in Fig.~\ref{fig:Delta_Ls}, the residuals $\Delta$ and $\varepsilon_i$
decay exponentially with $L_s$, with a clear hierarchy
$\Delta\gg\varepsilon_h\gg\varepsilon_3\equiv\varepsilon_5$. This is strong
evidence for the ultimate recovery of U($2N$) (the convergence rate is strongly
dependent on both $\beta$ and system volume), and moreover suggests the optimal
simulation strategy is to focus on the twisted condensate
$i\langle\bar\psi\gamma_3\psi\rangle$ for which  finite-$L_s$ corrections are
minimal. The equivalence of $\gamma_3$ and $\gamma_5$ condensates at finite
$L_s$ and their
superior convergence to the large-$L_s$ limit was shown analytically in
\cite{Hands:2015dyp}, where the convergence to fermions obeying 2+1$d$ Ginsparg-Wilson
relations was also demonstrated. Exponential improvement of convergence was 
shown in the large-$N$ limit of the GN model in \cite{Hands:2016foa}.  The benefits of a 
twisted mass term for improved recovery of U(2$N$) symmetry were also observed in
a study of non-compact QED$_3$ using Wilson fermions in \cite{Karthik:2015sgq}.

\section{The Thirring model with domain wall fermions}\label{sec:ThirDWF}

Even after settling on DWF, we still encounter some remaining
formulational issues. Just as for the staggered model, we have chosen a linear
interaction between the fermion current and a 2+1$d$ vector auxiliary field
$A_\mu$ defined on the lattice links. The simplest approach to formulating the Thirring model is
to restrict the interaction to the physical fields (\ref{eq:target}) defined on the domain walls at
$s=1,L_s$. This follows the treatment of the GN model with DWF developed in
\cite{Vranas:1999nx}. It has the technical advantage that the Pauli-Villars fields
required to cancel bulk mode contributions to the fermion
determinant do not couple to $A_\mu$, and hence can be safely excluded from the
simulation, which brings significant cost savings. In what follows this approach
will be referred to as the {\em Surface\/} model.

However, following the discussion below eq.~(\ref{eq:gt}) suggesting the strong
similarity of $A_\mu$ with an abelian gauge field, we also consider a {\em Bulk}
formulation in which $\Psi$, $\bar\Psi$ interact with a ``static''
field (ie. $\partial_3 A_\mu=0$) throughout the bulk:
\begin{equation}
{\cal S}=\bar\Psi{\cal D}\Psi=\bar\Psi D_W\Psi+\bar\Psi D_3\Psi+m_iS_i,
\label{eq:SDWF}
\end{equation}
with 
\begin{equation}
D_W=\gamma_\mu D_\mu-(\hat D^2+M);\;\;\;D_3=\gamma_3\partial_3-\hat\partial_3^2;
\end{equation}
and $m_iS_i$ is the explicit mass term defined only on the walls. Here 
\begin{eqnarray}
D_{\mu xy}&=&{1\over2}\left[(1+iA_{\mu x})\delta_{x+\hat\mu,y}-(1-iA_{\mu
x-\hat\mu})\delta_{x-\hat\mu,y}\right],\nonumber\\
\hat D^2_{xy}&=&{1\over2}\sum_\mu\left[(1+iA_{\mu x})\delta_{x+\hat\mu,y}+(1-iA_{\mu
x-\hat\mu})\delta_{x-\hat\mu,y}-2\delta_{xy}\right],
\end{eqnarray}  
and $M$ is the domain wall height, here set equal to 1. $\partial_3$, $\hat\partial_3^2$ are defined
similarly using finite differences in the 3 direction, respecting the open
boundary conditions, and with no coupling to
the auxiliary. These definitions imply the following properties:
\begin{equation}
[\partial_3,D_\mu]=[\partial_3,\hat D^2]=0\;\;\;\mbox{but}\;\;\;
[\partial_3,\hat\partial_3^2]\not=0\;\;\;\mbox{on domain walls.}
\label{eq:comms}
\end{equation}
The action (\ref{eq:SDWF}) may be simulated using the HMC algorithm for $N$
even; however the failure of the third commutator in (\ref{eq:comms}) to vanish
everywhere is an
obstruction  to proving $\mbox{det}{\cal D}$ is positive definite, so $N=1$ is
simulated using the RHMC algorithm with functional measure $\mbox{det}({\cal
D}^\dagger{\cal D})^{1\over2}$.

\section{Numerical results}

We have studied both surface and bulk formulations of the Thirring model in the
coupling range $ag^{-2}\in[0.1,1.0]$ with first $N=2$~\cite{Hands:2016foa} and now $N=1$. The RHMC
algorithm used 25 partial fractions to estimate fractional powers of ${\cal
D}^\dagger{\cal D}$. Most results are obtained on $12^3\times16$ ($N=2$) or
$12^3\times8$ ($N=1$). An exploration of volume and finite-$L_s$ effects for
$N=2$ was presented in \cite{Hands:2016foa}. Summary results for the bilinear condensate
$i\langle\bar\psi\gamma_3\psi\rangle$ with $m_3a=0.01$ are shown in
Fig.~\ref{fig:cond_N}. The bulk model shows
signifcantly enhanced pairing for $g^{-2}\lapprox0.5$ compared to the
surface model, and as might be anticipated pairing is greater for $N=1$
than $N=2$. This trend continues until a maximum at $ag^{-2}\approx0.2$. In
previous work with staggered fermions this has been identified with the effective location of the
continuum strong coupling limit~\cite{Christofi:2007ye}, 
since for stronger lattice couplings there is a breakdown
of reflection positivity~\cite{DelDebbio:1997dv}. We are thus confident that the range of couplings
explored includes the strong coupling limit. 

\begin{figure}
\begin{minipage}[t]{70mm}
\begin{center}
\includegraphics[width=7.5cm]{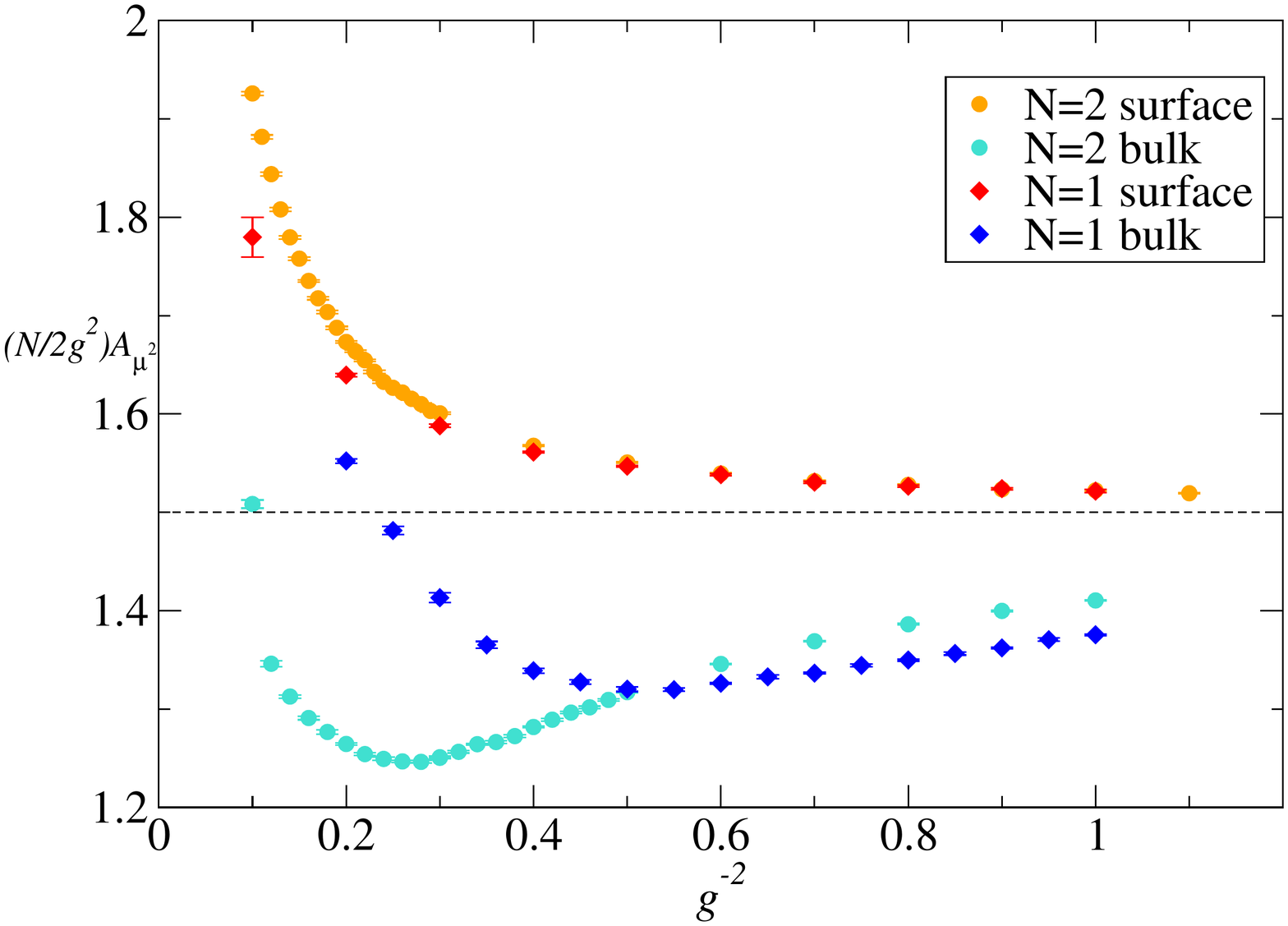}
\caption{Auxiliary action vs. $g^{-2}$.}
\label{fig:bose_N}
\end{center}
\end{minipage}
\hspace{\fill}
\begin{minipage}[t]{70mm}
\begin{center}
\includegraphics[width=7.5cm]{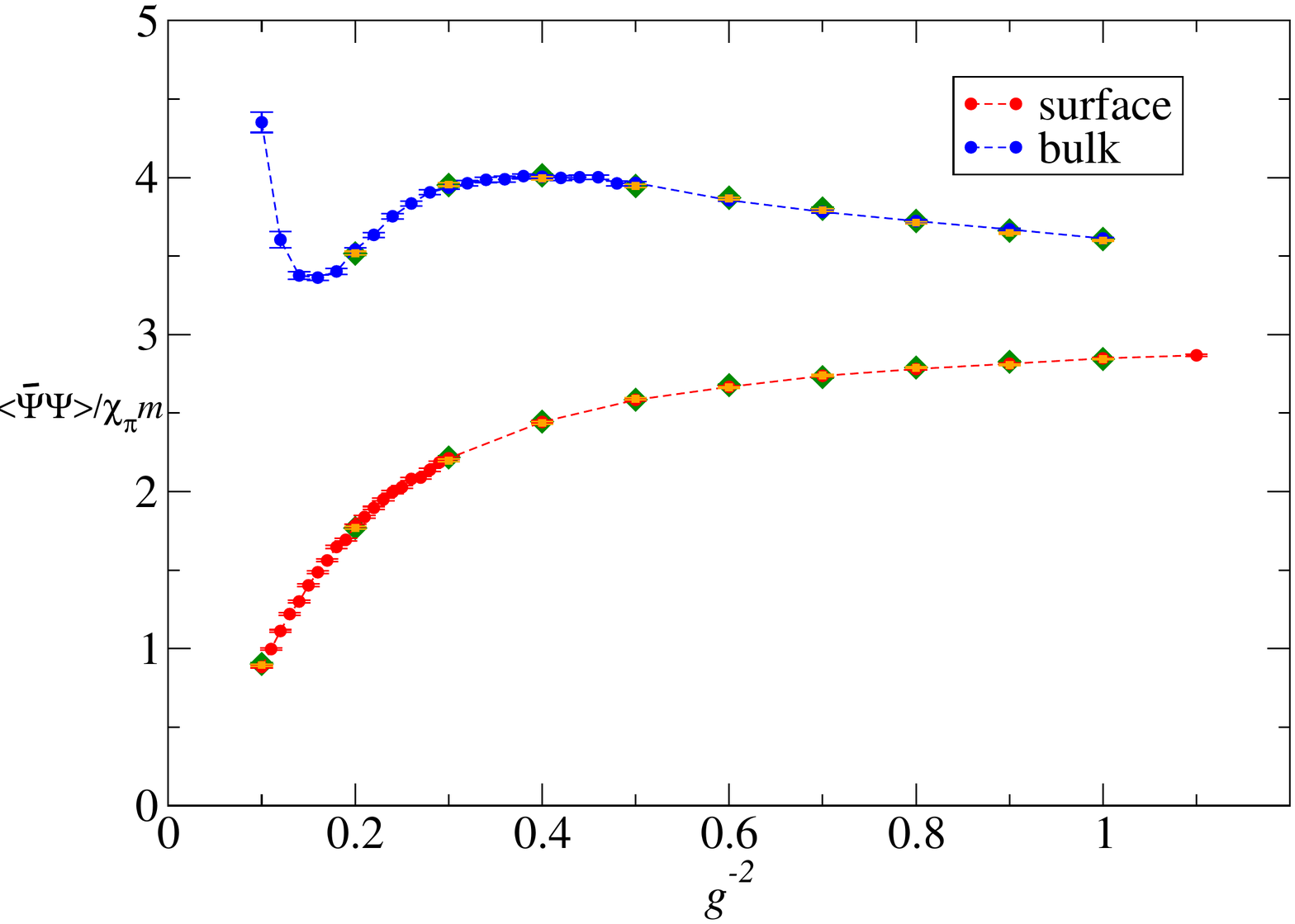}
\caption{Test of the axial Ward Identity.}
\label{fig:ratio}
\end{center}
\end{minipage}
\end{figure}

Fig.~\ref{fig:bose_N} shows the auxiliary action over the same coupling range, compared with the
free field value ${d\over2}$. The difference between surface and bulk models
is apparently very striking, but it should be borne in mind that this is really
a comparison of UV properties of two different regularisations of 
ostensibly the same continuum field theory.
Once again, there is tentative evidence for a change of behaviour of the $N=1$ bulk
model at $ag^{-2}\approx0.5$.

Fig.~\ref{fig:ratio} shows the ratio
$i\langle\bar\psi\gamma_3\psi\rangle/m_3\chi_\pi$ 
obtained for $N=2$, with the transverse susceptibility 
defined 
\begin{equation}
\chi_\pi=N\sum_x\langle\bar\psi\gamma_5\psi(0)\bar\psi\gamma_5\psi(x)\rangle.
\end{equation}
For a theory where the $\psi,\bar\psi$ fields respect U($2N$) symmetry, the
2+1$d$ generalisation of the axial Ward identity predicts the ratio
to be unity. Fig.~5 shows that this requirement is far from being met, and that
further work is needed to understand and calibrate the identification of the
physical fields via relations such as (\ref{eq:target}). Again, the disparity
between bulk and surface models is striking, with neither being obviously
preferred.

\begin{figure}
\begin{minipage}[t]{70mm}
\begin{center}
\includegraphics[width=6.5cm]{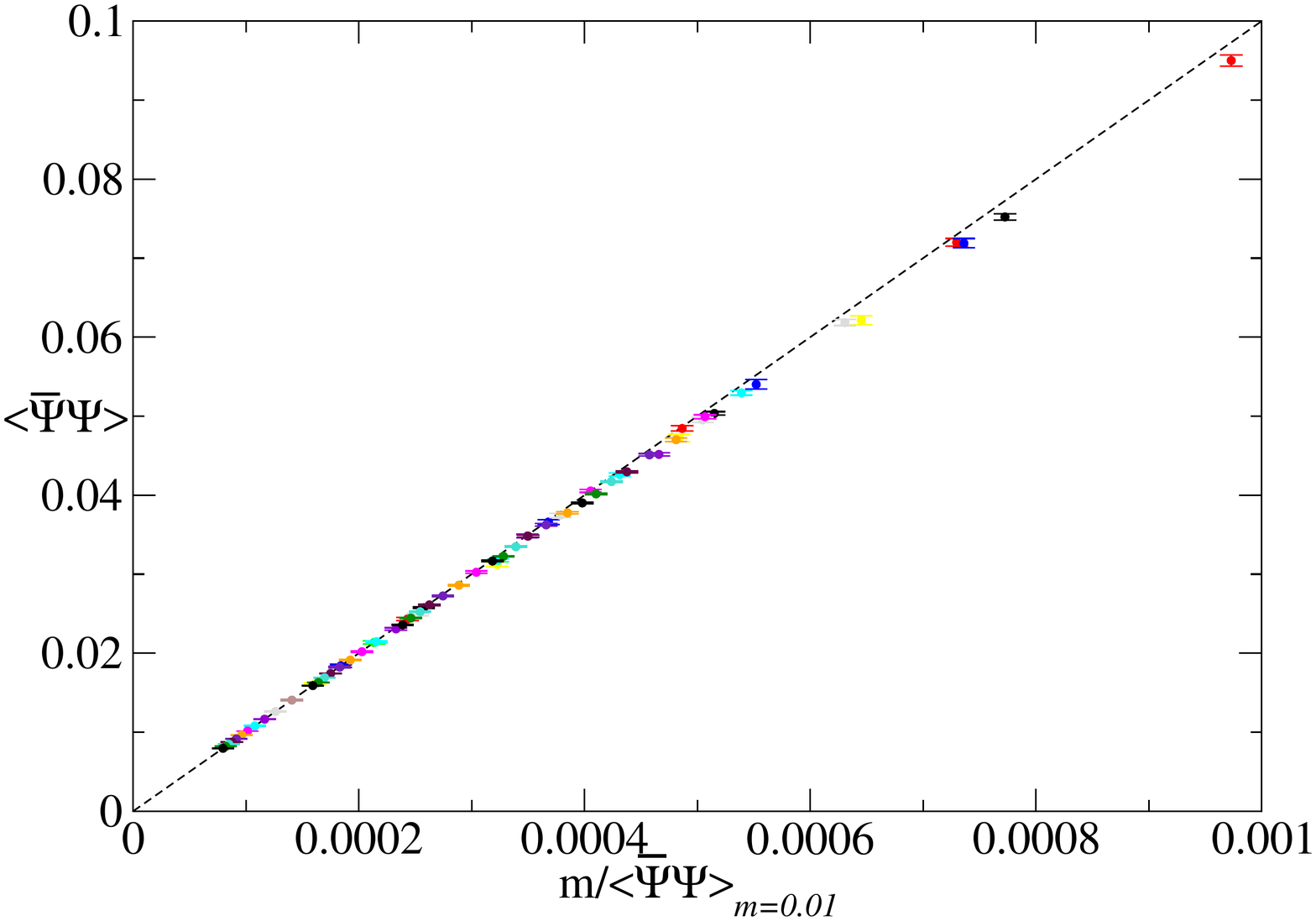}
\caption{Data collapse for the $N=1$ bulk model\dots}
\label{fig:collapse_bulk}
\end{center}
\end{minipage}
\hspace{\fill}
\begin{minipage}[t]{70mm}
\begin{center}
\includegraphics[width=6.5cm]{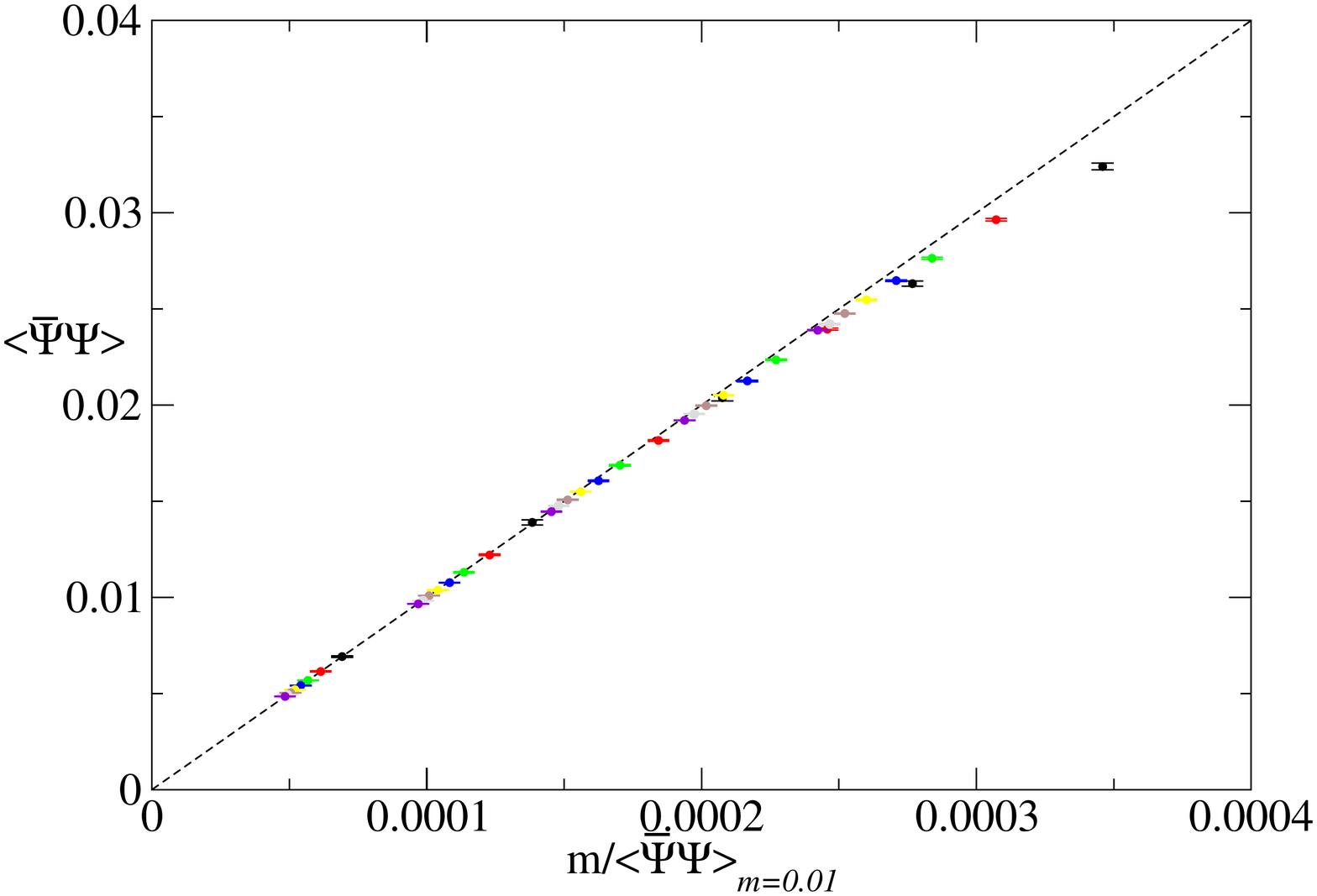}
\caption{\dots and for the surface model.}
\label{fig:collapse_surface}
\end{center}
\end{minipage}
\end{figure}
Finally we consider whether U($2N$) is spontaneously broken at strong coupling.
In \cite{Hands:2016foa}
the bilinear condensate for $N=2$ was examined as a function of bare mass $m$
across a range of couplings, and in every case a linear scaling
$\langle\bar\psi\psi\rangle\propto m$ was found, indicative of U($2N$) symmetry
being manifest as $m\to0$. We now extend this study to $N=1$.
Figs.~\ref{fig:collapse_bulk},\ref{fig:collapse_surface} show 
$\langle\bar\psi\psi(m)\rangle$ for $ma=0.01,\ldots,0.05$; with a trivial
rescaling implemented by choosing the abscissa as
$m/\langle\bar\psi\psi(am=0.01)\rangle$, data from the entire range of couplings
studied collapses onto a near linear curve, for both bulk (left) and surface
(right) models. There is no evidence for any singular behaviour associated with
a symmetry-breaking phase transition, and it seems safe to conclude
$\lim_{m\to0}\langle\bar\psi\psi\rangle=0$. Assuming this picture persists in the
large volume and $L_s\to\infty$ limits, this provides strong evidence that the
critical flavor number in the U($2N$)-symmetric Thirring model is constrained by
\begin{equation}
N_c<1.
\end{equation}

\section{Summary and outlook}

It has been shown that it is feasible to use DWF to study U($2N$)-symmetric
fermions in 2+1$d$, and that use of a twisted mass term $\sim
im_3\bar\psi\gamma_3\psi$ optimises the recovery of the symmetry as
$L_s\to\infty$. A study of the Thirring model at strong fermion self-couplings
then shows that DWF capture a very different physics to that described by
staggered fermions, which are governed by a different global symmetry away from
the weak-coupling long-wavelength limit. While it is still not possible to settle on the preferred
formulation of the strongly-coupled Thirring model, both the bulk and surface
versions presented here are in agreement that the critical flavor number
$N_c<1$. Fortunately this is compatible with the results obtained using a
distinct U($2N$)-symmetric approach involving the SLAC
derivative~\cite{Wellegehausen:2017goy}, also
presented at this conference~\cite{Jena}. Recent studies of U($2N$)-symmetric QED$_3$,
an asymptotically-free theory, have also concluded
$N_c<1$~\cite{Karthik:2015sgq,Karthik:2016ppr}. 
It is noteworthy that a disparity between DWF
and staggered fermions in a very different physical context, namely near a conformal
fixed point in a 3+1$d$ non-abelian gauge theory, has also been reported~\cite{Anna}.

In future work it will obviously necessary to study the effects of
$L_s\to\infty$, $V\to\infty$, and of varying the domain wall height $M$. It will
also be valuable to study the locality properies of the corresponding 2+1$d$
overlap operator, furnishing a non-trivial test of the DWF approach in a new
physical context. Another question to ponder is what does chiral symmetry
breaking actually look like for 2+1+1$d$ DWF? We are currently investigating
this issue by quenching the Thirring model. Finally, functional renormalisation
group studies indicate that in the hunt for a QCP it may be interesting to
include a U($2N$)-invariant Haldane interaction 
$-\tilde g^2(\bar\psi\gamma_3\gamma_5\psi)^2$~\cite{Gehring:2015vja}; the control offered by DWF will make
this a straightforward exercise.

\clearpage
\bibliography{granada}

\end{document}